\documentclass[%
reprint,
 amsmath,amssymb,
 aps,
]{revtex4-1}

\usepackage{graphicx}% Include figure files
\usepackage{dcolumn}% Align table columns on decimal point
\usepackage{bm}% bold math
\def\dd{{\rm d}}\def\_#1{^{}_{#1}}
\def\half{{\textstyle\frac12}}
\def\beq{\begin{equation}}\def\eeq{\end{equation}}
\def\bea{\begin{eqnarray}}\def\eea{\end{eqnarray}}
\usepackage[bottom]{footmisc}
\usepackage{natbib}
\usepackage{xcolor}
\newcommand{\bolen}[1]{{\color{red}[\textbf{Brett:} #1]}}
\newcommand{\shane}[1]{{\color{blue}[\textbf{shane:} #1]}}

\usepackage{hyperref}
\hypersetup{
    colorlinks=true,     % false: boxed links; true: colored links
    linkcolor=red,       % color of internal links
    citecolor=blue,      % color of links to bibliography
    filecolor=blue,      % color of file links
    urlcolor=blue        % color of external links
}

\begin{document}

%% Understanding exotic black hole orbits using  undergraduate mechanics
\title{Understanding exotic black hole orbits using effective potentials}

\author{Steven Pakiela} \affiliation{Department of Physics,
Grand Valley State University, Allendale, MI 49504\\ \kern10pt}
\author{Monica Rizzo}\affiliation{Center for Interdisciplinary Exploration and Research in Astrophysics, Northwestern University, Evanston, IL 60208, USA}
\author{Brett Bolen}\email{bolenbr@gvsu.edu}
\affiliation{Department of Physics,
Grand Valley State University, Allendale, MI 49504\\ \kern10pt}
\author{Benjamin P. Holder}\email{holderbe@gvsu.edu}
\affiliation{Department of Physics,
Grand Valley State University, Allendale, MI 49504\\ \kern10pt}

\author{Shane L. Larson}\email{s.larson@northwestern.edu}\affiliation{Center for Interdisciplinary Exploration and Research in Astrophysics, Northwestern University, Evanston, IL 60208, USA}

\preprint{APS/123-QED}

\date{\today}% It is always \today, today,

\begin{abstract}
Gravitational wave astronomy is an emerging observational discipline that expands the astrophysical messengers astronomers can use to probe cosmic phenomena. The gravitational waveform from a source encodes the astrophysical properties and the dynamical motion of mass within the system; this is particularly evident in the case of binaries, where the overall amplitude of the system scales with physical parameters like mass and distance, but the phase structure of the waveform encodes the orbital evolution of the system. Strongly gravitating systems can show interesting and unusual orbital trajectories, as is the case for ``extreme mass ratio inspirals,'' observable in the millihertz gravitational wave band by space-based gravitational wave detectors. These sources can exhibit ``zoom-whirl'' orbits, which make complicated waveforms that are useful for mapping out the gravitational structure of the system. Zoom-whirl behavior can be intuitively understood in the context of effective potentials, which should be familiar to students from classical orbital theory in mechanics. Here we demonstrate and explain zoom-whirl orbits using effective potential theory around Schwarzschild black holes, and present an \href{https://ciera.northwestern.edu/gallery/zoom-whirl/}{interactive tool} that can be used in classroom and other pedagogical settings.
\end{abstract}

\pacs{Valid PACS appear here}% PACS, the Physics and Astronomy
\maketitle

%\tableofcontents
%\section{Tentative outline}
%This is a tentative outline of our AJP paper (will not appear in article)
%\begin{enumerate}
%\item EMRI as a major LISA source. Rah,rah
%\item Effective potential for Schwarzschild
%\begin{enumerate}
%\item Newtonian vs Schwarzschild effective potential
%\item discuss marginally bound orbits 
%\item how when one approaches a turning point that even though the total energy is constant because there are now three turning point the change in kinetic radial energy is small so all the change in energy is in angular kinetic energy.  Thus we can explain the ``whirling'' behavior.
%\end{enumerate}
%\item what is ``whirling''

%\item extension into Kerr
%\end{enumerate}

%% ================ INTRODUCTION ==================================
%% ================================================================
\section{Introduction}
Traditional training in physics revolves around developing core paradigms that underpin much of the discipline. Some ideas, such as waves and conservation laws, are quite general and applicable across many different areas of research, an observation that is quite useful pedagogically \cite{paradigms}. % https://doi.org/10.1119/1.1374248

    Such paradigms are not only useful for solving problems, but also for understanding seemingly complex and exotic behavior in physical systems. This paper illustrates how effective potentials, typically introduced to students in the context of classical mechanics and scattering theory, can be utilized in gravitational wave astronomy to understand ``zoom-whirl'' orbits. These orbits are expected to be common in extreme-mass-ratio-inspirals (EMRIs) \cite{CKP} of small compact objects into supermassive black holes. These are a key source for the Laser Interferometer Space Antenna (LISA), expected to launch in the early 2030s \cite{NASA_LISA}. Zoom-whirl orbits have complex structure that when visualized appear random and chaotic, however studies of the dynamical behaviour of these orbits have shown they are not formally chaotic \cite{Brinicki}. The orbits  fill the volume around the central black hole, a behavior that is one of the reasons EMRIs are valuable observational sources: the trajectory is encoded in the gravitational wave signal and if it can be reconstructed, provides a detailed map of the spacetime around the black hole. This is completely analogous to \textit{geodesey}, the careful mapping of the gravitational environment of the Earth from observations of satellite trajectories.

This paper is organized as follows: in Section \ref{sec.emri} we review the basics of EMRIs and effective potentials; in Section \ref{sec.Schwarzschild} we provide the effective potential framework for orbits around a Schwarzschild black hole; in Section \ref{sub.whilringPotential} we apply the effective potential theory to the Schwarzschild case and illustrate how zoom-whirl orbits arise. In Section \ref{sec.gravWaves}, we give a brief overview of both the production and detection of gravitational radiation. In section \ref{sec.tool}, we discuss on online visualiztion tool and how it might be used in the undergraduate classroom. Lastly, in Section \ref{sec.discussion} we discuss extension of this method to more complicated cases, and discuss the utility of this approach in the classroom.

Throughout the paper we have used geometricized units, which are commonly used in gravitational wave physics. In these units $G = c = 1$, and the dimensionality of all quantities is in powers of length. In particular, time is measured in meters, mass is measured in meters, energy is measured in meters, and angular momentum is measured in meters squared. Note that throughout the paper, specific energy (energy per unit mass) is used, so is dimensionless in geometricized units; similarly specific angular momentum (angular momentum per unit mass) is used, so has units of meters in these units. To restore conventional SI units to any formula, multiply masses by a factor of $G/c^{2}$, energies by a factor of $G/c^{4}$, and angular momenta by a factor of $G/c^3$. 

The codes used to develop the graphics have been made publicly available as python scripts and Jupyter notebooks, as have animations of the still-frame graphics provided in this paper.\cite{zoomWhirlGithub}

%% ======= Section : EMRIs ========================================
%% ================================================================
\section{Extreme Mass Ratio Inspirals}\label{sec.emri}

Bertrand's theorem  \footnote{An English translation of Bertrand's original paper ]url{https://arxiv.org/abs/0704.2396}} states that the only central force laws which give rise to closed orbits for all bound particles are the inverse square law and Hooke's law.  Thus, orbits in Einstein's theory of general relativity cannot be closed.  In most situations, it is conventional to treat solar system dynamics with Newtonian gravity, as it provides reasonable accuracy for most calculations. However precision measurements, such as the observations of Mercury's perihelion precession, reveal the underlying fact that Newtonian gravity is simply the weak field limit of general relativity. As will be shown in Section \ref{sec.effPotential}, this manifests itself clearly in the form of the effective potential, which shows the implication of Bertrand's theorem on particle orbits.

\begin{figure}[h]
\includegraphics[width=
\columnwidth]{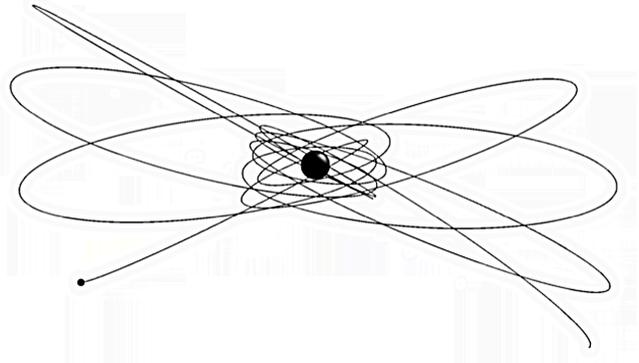}
\caption{A classic zoom-whirl trajectory around a spinning black hole (\cite{DrascoPrivate}).}.\label{fig.drascoZoomWhirl}
 \end{figure}

Zoom-whirl orbits in extreme mass ratio systems are another example where Bertrand's theorem manifests itself. Compact stellar remnants, such as neutron stars or stellar-mass black holes of mass $\sim 10\, M_\odot$, are expected to be captured by massive black holes of mass $\gtrsim 10^6\, M_\odot$  in the centers of galaxies \cite{2007CQGra..24R.113A}. The orbits are expected to be highly eccentric, where the light companion ``zooms'' inward from large radii, and makes a close encounter with the central black hole. The orbit will show a degree of perihelion precession during the close encounter, but under certain conditions the precession can be extreme, resulting in a ``whirl'' where the companion makes many loops around the central mass before zooming out to large radii again. Over time, the emission of gravitational-radiation extracts energy from the orbit, causing it to shrink and become more and more circular until the compact body plunges into the central black hole. A three dimensional simulation of such a trajectory around a spinning (Kerr) black hole is shown in Figure \ref{fig.drascoZoomWhirl}.

In astrophysical contexts, the central black holes are expected to be rotating black holes represented by the Kerr solution.  As we show here the zoom-whirl behavior is not a consequence of the spin of the black hole, but rather a feature of the potential the compact particle is moving in. To that end, we will derive the zoom-whirl behavior from equations of motion of a test particle in the fixed spacetime of a stationary Schwarzschild (spherical, non-spinning) black hole.

%% ======= Section : Effective Potentials =========================
%% ================================================================
\section{Effective Potentials} \label{sec.effPotential}

\subsection{Comparison of Newtonian and Schwarzschild effective potentials}\label{sec.Schwarzschild}

From classical mechanics, the equation of motion of a particle in orbit around a massive body has the form
\beq{} \label{EOM}
m \ddot r =- \frac{\dd}{\dd r} \left(V(r) + \frac{1}{2} \frac{\ell^2}{m r^2} \right)
\eeq
where    $\ell= r^2 \dot \phi$ is a constant of the motion representing angular momentum and an over-dot represents a time derivative. The energy $(E)$ is also a conserved quantity.
In Newtonian gravity, $V(r)=M m/r$, and the term inside the parenthesis in Eq.\ \ref{EOM} is called the ``effective potential.'' Normalizing the conserved quantities by the mass of the orbiting particle generalizes this to the specific effective potential $(\tilde \ell =\ell/m\, \textrm{and}\, \tilde V=V/m)$  with form
\beq{}
\tilde V_{\substack{\textrm{Newton} \\ \textrm{Effective}}} (r) = -\frac{  M }{r} + \frac{\tilde \ell^2}{2  r^2} \ .
\eeq 

\begin{figure}[h]
\includegraphics[width=
\columnwidth]{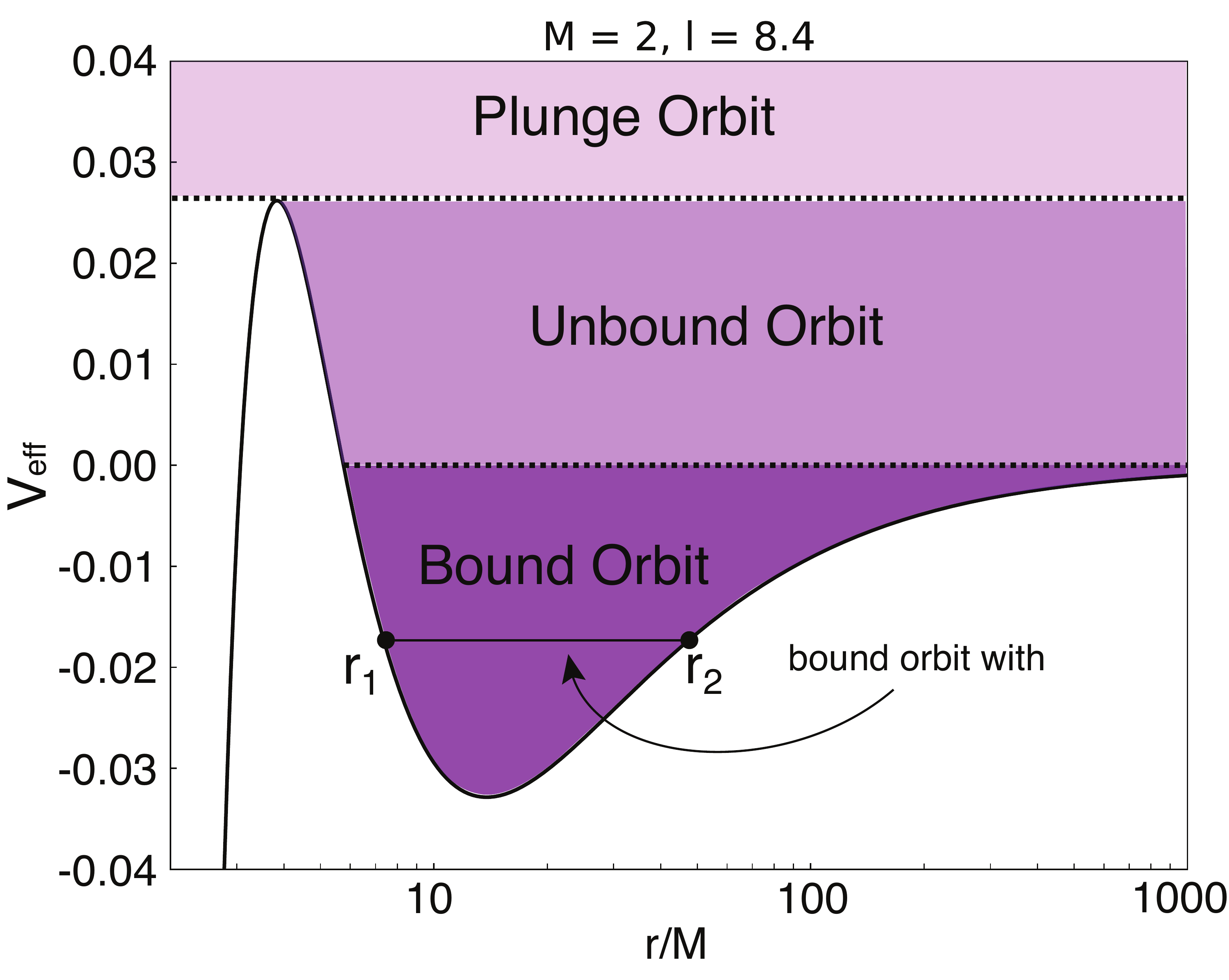}
\caption{The effective potential for a Schwarzschild black hole, showing the regimes for bound and  unbound orbits.}.
\label{fig.VeffOrbitsBinding}
 \end{figure}

The effective potential is useful for classifying orbital geometries by were the specific orbital energy $(\tilde E)$ exists relative to the effective potential; there are four general cases, graphically shown in Figure~ \ref{fig.VeffOrbitsBinding}.  Circular orbits at constant radii occur at the global minimum of the potential, where $\dd \tilde V/\dd r =0$. For bound orbits, $\tilde E < 0$ but larger than minimum of the potential: these orbits are elliptical with inner (perapasis) and outer (apoapsis) turning points, defined by the radii where the energy ${\tilde E}$ is equal to the effective potential ${\tilde V}$:
\beq{}
r_\pm= - \frac{M}{2 \tilde E} \left( 1 \pm \sqrt{\frac{2 \tilde E \ell^2}{M^2} +1} \right)  \, . \label{newTP}
\eeq
If the energy is exactly zero, the orbits are parabolic, and if ${\tilde E}>0$, then the orbits are hyperbolic.

\begin{figure}
    \centering
   \includegraphics[width=1 \columnwidth]{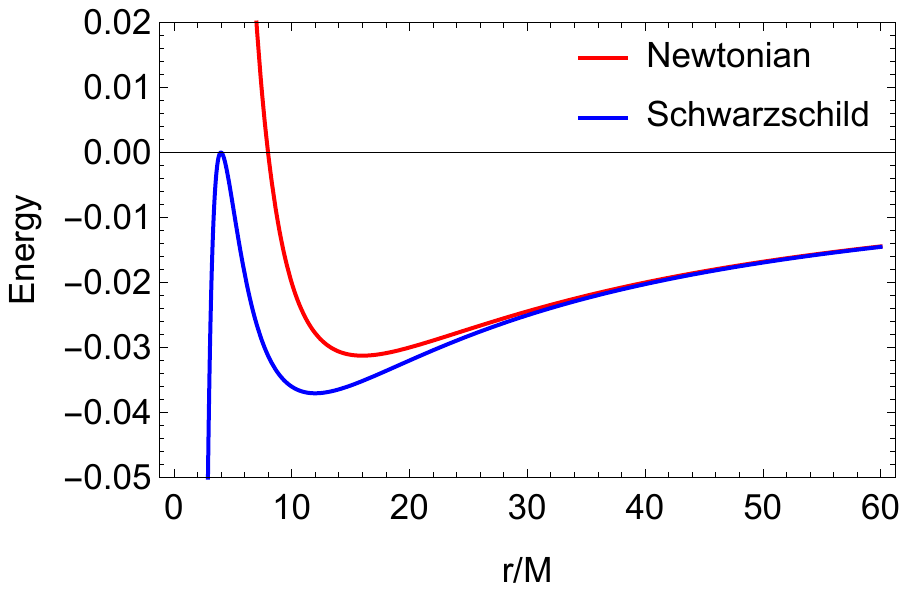}
   \caption{Comparison of the classic Newtonian effective potential and the Schwarzschild effective potential where $\ell=4$ and $M=1$. Note how the additional $1/r^3$ term in the Schwarzschild case dominates at small radii, creating a local maximum.}
   \label{fig.VeffSchwarzschild}
\end{figure}

In general relativity one can derive an equation of motion for a Schwarzschild (non-rotating) black hole  that is similar to Eq.\ \ref{EOM}, and cast it in the same language of effective potentials. Energy and angular momentum are again conserved, but the effective potential has an additional term:
\footnote{The Schwarzschild solution is a spherically symmetric solution to the Einstein equation.  It represents a static black hole. A full derivation of this equation can be found in any undergraduate text such as Hartle or Moore}
\bea{}
\tilde V_{\substack{\textrm{Schwarzschild} \\ \textrm{Effective}}} &=& - \frac{ M}{r} + \frac{\tilde{\ell}^2}{2 r^2} - \frac{M \tilde{\ell}^2}{ \,r^3}\nonumber \\
&=&-\frac{ M}{r} + \frac{\tilde{\ell}^2}{2 r^2}
\left(1 - \frac{R_s}{r} \right)  
\eea 
Here $R_s=2 M$ is the Schwarzschild radius.  The third term scales like $1/r^{3}$; it is negligible at large radii giving behaviour that looks like classic Newtonian orbits (as expected), but dominates at small radii near the central black hole.  This potential is illustrated in Figure \ref{fig.VeffSchwarzschild} together with the classic Newtonian central potential. The effect of the dominant term near the black hole creates a second extremum in the potential, with a local maximum near the black hole. As in the Newtonian case, the overall shape and structure of this potential makes classification of orbital behaviors possible based on the relationship of the energy of the orbit to the value of the effective potential. 

\subsection{ Whirling in Effective Potentials}\label{sub.whilringPotential}
The difference between the Newtonian and Schwarzschild effective potentials is responsible for the famous perihelion advancement observed in Mercury's orbit; explaining the anomalous advance and was one of the first proofs of Einstein's general relativity. \cite{nodoubt}  

To see the origin of the perihelion precession from the effective potential, note the angular speed $u_{\phi}$ must maintain a value consistent with the conservation of angular momentum. Since 
\beq 
u^{\phi}= \frac{\tilde \ell}{r^2}
\label{eqn.AngSpeed}
\eeq 
as $r$ becomes smaller, the value of $u^{\phi}$ increases. In Figure \ref{fig.VeffSchwarzschild}, the inner turning point in the Schwarzschild potential is \textit{always} at smaller radii than in the Newtonian potential, meaning the angular speed is higher as the particle approaches the turning point, and more orbital phase is accumulated, resulting in an anomalous advance in the periapsis on the next orbit.

The unique shape of a black hole's effective potential is also responsible for the whirling behavior in extreme mass ratio systems. Roughly speaking, the radial kinetic energy scales as the difference between the energy of the orbit, $\tilde{E}$, and the effective potential, $\tilde{V}_{eff}$. Thus ${\dot r}$ changes rapidly when $\tilde{V}_{eff}$ changes rapidly, and ${\dot r}$ changes slowly when $\tilde{V}_{eff}$ changes slowly. As the particle approaches the turning point of an orbit, ${\dot r} \rightarrow 0$. 

\begin{figure}
    \centering
   \includegraphics[width=\columnwidth]{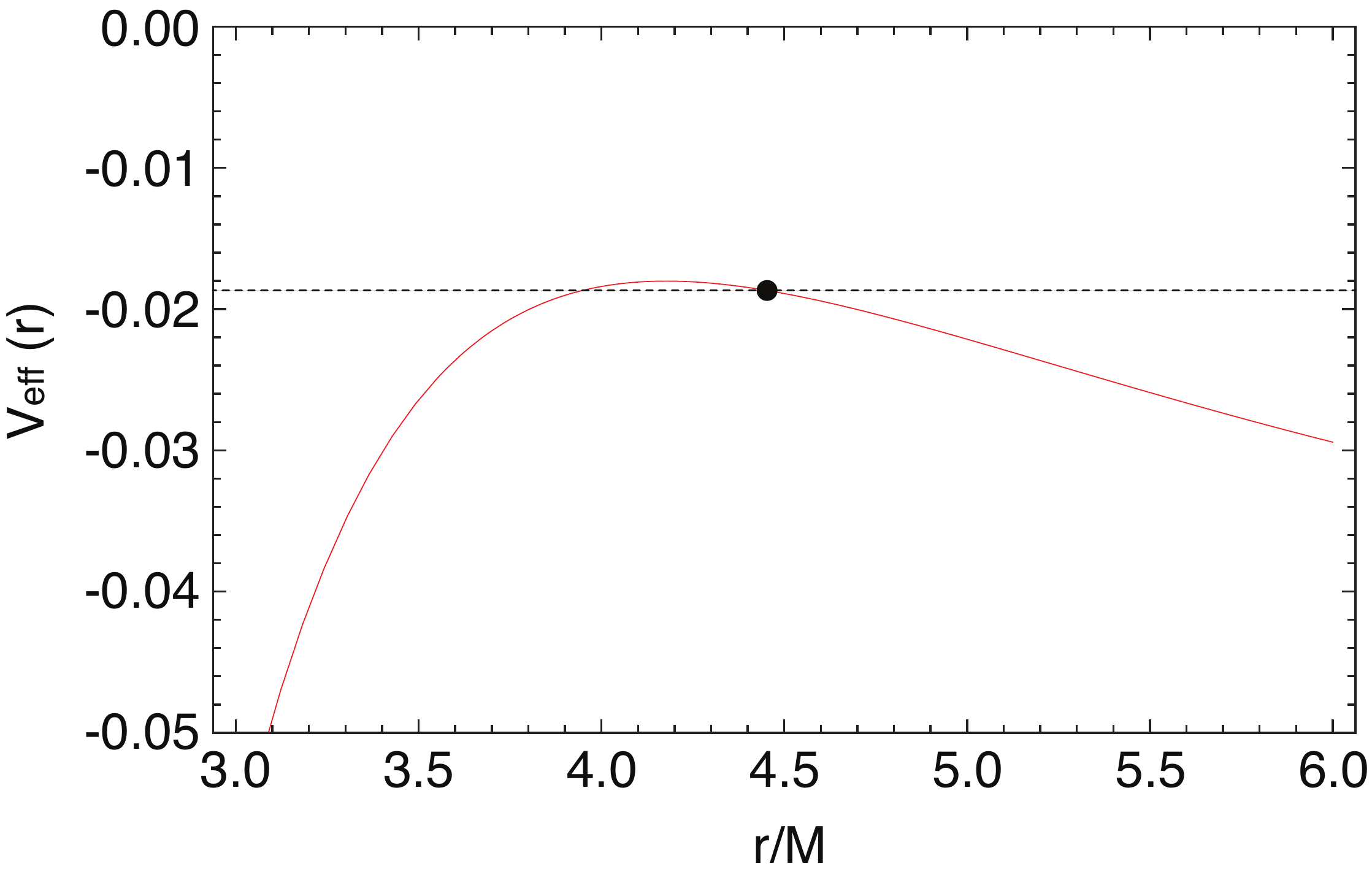}
   \caption{Zoom-in near the inner peak of the effective potential, plotted vs. $R$ in units of the black hole mass. The dashed line is the energy of the orbit, and the black dot indicates the inner turning point, where the particle energy matches the effective potential.}
   \label{fig.turningPeak}
\end{figure}

Consider a particle orbit whose energy $\tilde{E}$ is very close to the maximum energy at the inner peak of the Schwarzschild effective potential, such that the turning point (where $\tilde{E}$ crosses $\tilde{V}_{eff}$) is very near the extremum of the potential, as shown in Figure \ref{fig.turningPeak}. The turning point in this case is very near the extremum. When the particle is very close to the turning point, its radial speed ${\dot r}$ is very small but changing (relatively) slowly because the slope of the potential is shallow. At the same time, since the radius is approximately constant during the extended time the particle is in this region of the potential, $u^{\phi}$ is approximately constant and phase $\Delta \phi$ is rapidly accumulating -- the particle is ``whirling.'' Eventually the particle reaches its turning point, ${\dot r} = 0$, and begins moving toward larger radii in the potential; $u^{\phi}$ begins to decreases, and the accumulation of phase slows. The whirling phase ends as the particle moves to larger radii, and the zoom phase of the orbit commences again. This whirling is extreme precession, enabled by the presence of a local maximum in the potential.

\subsection{Explicit calculation of whirliness}\label{sub.mathWhirly}
%\subsection{More detail on whirl zoom} \label{sec.ZoomWhirl}

Periapsis precession is most often reported as an angle or a rate of change in an angular position; for instance, Mercury's perihelion advance is usually reported as 574 arcseconds/century. In the case of zoom-whirl orbits, the precession is enormous in a very short time, and it is convenient to define the ``whirliness'' of the orbit as
\begin{equation}
    \varpi = \frac{\Delta \phi}{2 \pi}
    \label{eqn.whirlyness}
\end{equation}
where $\Delta \phi$ is the total accumulated angular phase of the particle along its orbit between successive periapsis passes, as viewed by an observer far from the black hole. The normalization $2 \pi$ is the expected accumulated phase in a normal Keplerian orbit, so one can view the whirliness $\varpi$ as being the number of loops around the black hole a distant observer sees the particle make.

The computation of the whirliness $\varpi$ is most easily accomplished by writing out the components of the particle 4-velocity, $u^{\alpha}$, for the $r$ and $\phi$ components:
\begin{equation}
    u^{r} = {\dot r} = \sqrt{2 {\tilde E} - \frac{{\tilde \ell}^{2}}{r^{2}} + \frac{2 M}{r}}
    \label{eqn.GeodesicRdot}
\end{equation}
and
\begin{equation}
    u^{\phi}= \frac{\tilde \ell}{r^2}\ .
    \label{eqn.GeodesicPhidot}
\end{equation}

Depending on one's inclinations (or the skills of your students), one could use these equations to either numerically evaluate the whirliness, or construct a mathematical transformation that allows the whirliness to be written in closed form. 

% omit  Numerical integration is straight-forward: \bolen{Need to finish}

%************************ 
%
%There are two ways to do this section
%\begin{itemize}
%    \item In the talk  and his mathematica notebook shane's talk where he simply numerically integrates $\dd \phi/\dd r$, obtains a very nasty term involving elliptic integrals \shane{The only reason to integrate $d\phi/dr$ is to get the number of whirls. Our basic argument is that this becomes large near the peak of the effective potential (lots of dphi accumlates over small dr), which can be seen without integrating}
%    \item  A cleaner method is to use Possion and Will.  They reperameterize the integral in term of $\chi$ which replaces $\phi$ as $\chi$ goes from $0$ to $2 \pi$ in the time it takes to go from apoapsis to apoapsis. They then can calculate the perihelion advancement analytically.  It is this approach that I will try to outline here  
%\end{itemize}
%**********************************

A useful analytic form of the whirliness can be found by reexpressing the forms of ${\dot r}$ and ${\dot \phi}$ in terms of a new auxillary parameter. This is a classic method from the literature \citep{1959RSPSA.249..180D,1961RSPSA.263...39D,1994PhRvD..50.3816C}, but a modern textbook implementation can be found \textit{in extenso} in \cite{2014grav.book.....P}. Here we take the traditional approach, which describes the orbit in terms of the \textit{semi-latus rectum} ($p$) of an ellipse; geometrically this is hard to understand, particularly in the context of a highly precessing orbit, so one may favor rewriting the expressions in terms of the semi-major axis $a$ which appears in Kepler's third law, and thus related to an observable (period, $P$). The semi-major axis and semi-latus rectum are related by the eccentricity:
\begin{equation}
    p = a(1 - e^{2})\ .
    \label{eqn.slrDefined}
\end{equation}
The cost of favoring $a$ over $p$ will be bulky terms that depend on eccentricity; from a computational viewpoint these are constant for an orbit. Note also that the turning points of the orbit, the periapsis $r_{p}$ and apoapsis $r_{a}$ may also be written in terms of $a$ and $e$:
\beq{}
r_{p} = a (1-e) \, , \quad
r_{a} = a (1+e)\ .
% r_{p} =\frac{p}{1+e} \, , \quad
% r_{a} =\frac{p}{1-e}
\eeq 
Here $e$ is the relativistic eccentricity, which is $e=0$ for a circular orbit, and $0<e<1$ for a bound orbit, completely analogous to the Keplerian case.

The specific angular momentum and specific mechanical energy are
%\bea
%5&\tilde \ell^2 &= \frac{M a (1 - e^{2})}{1- \frac{1}{3} (R_s/a)\frac{\left(3+ e^2 \right)}{(1 - e^{2})}} \\
%&\tilde E& = - \frac{M}{2 a} \left[\frac{1 -2 (R_s/a)/(1 - e^{2})}{1 - \half (R_s/a)\frac{\left(3+ e^2 \right)}{(1 - e^{2})}}\right] \, ,
%\eea 
% v1 Eq. in terms of rp
%\bea
%&\tilde \ell^2 &= \frac{M a (1 - e^{2})}{1- \frac{1}{3} \frac{\left(3+ e^2 \right)}{(1 - e^{2})} R_s/a} \\
%&\tilde E& = - \frac{M}{2 a} \frac{(1 - e^{2}) -2 R_s/a}{(1 - e^{2}) - \half \left( 3+e^2\right) R_s/a} \, .
%\eea 
% Original Eq. in terms of SLR
\bea
&\tilde \ell^2 &= \frac{M p}{1- \frac{1}{3} \left(3+ e^2 \right) R_s/p} \\
&\tilde E& = - \frac{M}{2 p} \left(1 -e^2 \right) \frac{1 -2 R_s/p}{1- \half \left( 3+e^2\right) R_s/p} \, .
\eea 
where $R_{s} = 2M$ is the Schwarzschild radius. Note that when the semi-latus rectum is much larger than a Schwarzshild radius $(p\gg R_s)$, these equations give the Newtonian limit for elliptical orbits  ($\tilde \ell= \sqrt{M p}$ and $\tilde E= - (M/2p) (1-e^2)$).  

\begin{figure}
    \centering
   \includegraphics[width=1.0 \columnwidth]{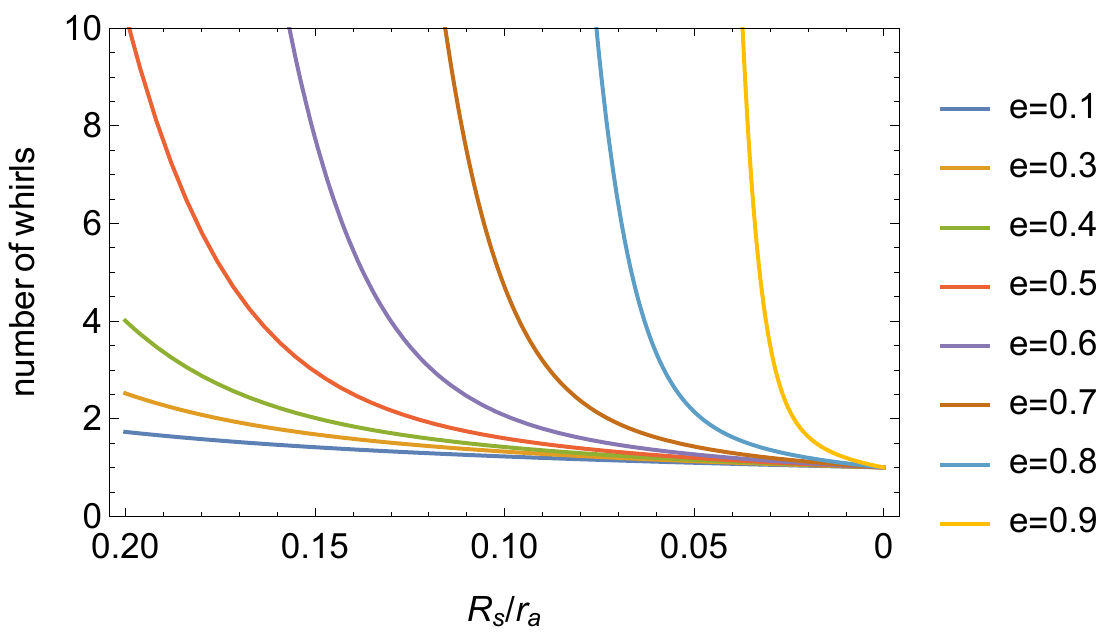}
   \label{fig.num_whirls}
    \caption{A plot of the “whirliness” of an orbit (the angular
    distance a test mass travels in one orbit divided by $2 \pi$) verses the ratio of the Schwarzschild radius to the apoapsis ($r_a$) for various values of the eccentricity. Note that there are two
    factors which govern how an orbit’s angular distance deviates from that of a Newtonian closed orbit. The first is that as eccentricity increases the closer the orbit’s apoapses approaches the inner unstable point $r_{IUCO}$ of the effective potential and the second is how close $r_a$ approaches the Schwarzschild radius. Both these facts are illustrated in the plot above and both terms may be understood by examining where the position of orbital energy on the effective potential plot.}
\end{figure}

Conventionally, a complete orbital period is defined as when the test particle has traveled from periapsis to periapsis in the coordinate $r$. For a Newtonian orbit the test particle has traveled an angle of in $\phi$ of $ 2 \pi$ during that period.  However, for precessing orbits, and zoom-whirl orbits in particular, in the time between two successive periapse passes in the coordinate $r$, it may have advanced in $\phi$ by more than $2 \pi$, a fact quantitatively captured in the whirliness, $\varpi$ as shown in Figure~\ref{fig.num_whirls}.

To derive a closed form expression for the whirliness, $\varpi$, it is convenient to parameterize the orbit in terms of an angular parameter $\chi$, rather than the conventional azimuthal angle $\phi$. This parameter has the value $\chi=0$ at periapsis and $\chi= \pi$ at apopsis; it is related to the normal azimuthal angle $\phi$ by
\beq
    \phi = \frac{1}{\sqrt{1 - 3 R_{s}/p}}\chi\ .
\eeq
This parameterization is convenient because it accounts for the curvature of space near the black hole horizon; at semi-latus rectum values far from the horizon, $\chi \approx \phi$, giving the classic Keplerian result. The shape equation of the orbit may be written in terms of this parameter as
 \beq
 r(\chi) = \frac{p}{1 + e \cos(\chi)}\ .
 \eeq
As expected, far from the black hole where $\chi \sim \phi$, this gives the shape equation in Newtonian gravity, $r(\phi)= p/(1+ e \cos \phi)$. The proposed definition of ``whirliness'' is asking how much angle $\chi$ the orbit evolves through compared to the Newtonian result of $\Delta \phi = 2\pi$ for a single orbit. Thus whirliness can be evaluated by finding $\dd \phi / \dd \chi$.
 
 Using the definitions of $\tilde \ell$ and $\tilde E$, the difference between the specific total mechanical energy and the specific potential energy is
 \beq{}
 \tilde E -\tilde V= \frac{M}{2 p} e^2 \sin^2 \chi
 \left( \frac{1-(3+ e \cos \chi) R_s/p}
 {1-\half(3+e^2)R_s/p}\right)\ .
 \eeq
 \noindent From energy conservation, the total specific mechanical energy is
 \beq{}
 \frac{1}{2} \dot r^2 +\tilde V = \tilde{E}\ .
 \eeq
Solving this for $\dot r=\dd r/\dd \tau=\sqrt{2(\tilde E-\tilde V)}$, and then using the inverse of this relationship to find $\dd \phi/\dd \chi= \dot \phi (\dd \tau/ \dd \chi)$ yields
 \beq \label{angrelate2}
 \frac{\dd \phi}{ \dd \chi} = \frac{1}{\sqrt{1- \left( 3+ e \cos \chi\right) R_s/p}} \, .
 \eeq 
Thus in the Newtonian limit $(p \gg R_s)$, the two angles, $\phi$ and $\chi$, are the same as one would expect.  Furthermore, by treating the term $R_s/p$ as small and Taylor expanding in this term, Equation~\ref{angrelate2}  can be integrated to give 
\bea
\Delta \phi &=& \int_0^{2\pi} \frac{\dd \phi}{\dd \chi} \dd \chi \nonumber \\
&=& 2 \pi + 6 \pi \left(\frac{R_s}{ p} \right) +
\frac{3 \pi}{2}(18+e^2) \left(\frac{R_s}{ p} \right)^2  + \ldots
\label{eqn.periapsisExpansion}
\eea 
Here, the leading term is the classic Keplerian result for a closed orbit, and the difference between the leading term and the first order term is the classic periastron shift term for Schwarzschild. To leading order, $\Delta \phi - 2\pi$ gives the shift in periapsis from pure Newtonian gravity and $(\Delta \phi/(2 \pi) -1 )$ is the excess whirliness.

Here we have been talking about whirliness in the context of eccentric orbits, with $e \neq 0$. However, examination of Equation~\ref{eqn.periapsisExpansion} shows that the first order term does not depend on eccentricity, so does not tend to zero when the orbit becomes circular. What does it mean to have a periastron shift for a circular orbit? In this case, this is a manifestation of the curved structure of spacetime in Schwarzschild.

In the limit where the semi-latus rectum is close to the Schwarzschild radius $(p \approx R)$, one finds the so-called ``whirl zoom" orbits where the orbiting particle has essentially two orbital scales.  First is the zoom phase where the orbit is quickly going around very close to the primary mass and a second whirl phase where the orbiting body goes far away from the primary mass.

%% ==============================
\section{Gravitational Waves}\label{sec.gravWaves}

In electromagnetism, accelerating electric charges give rise to the system radiating electromagnetic radiation. The analogous phenomena happens when masses are accelerated, giving gravitational radiation in general relativity. In astrophysical scenarios, particles are always accelerating along their orbit, so the emitted gravitational radiation encodes the properties of the orbit. 

The interest in zoom-whirl orbits around massive black holes is motivated by the fact that these are prospective sources for low-frequency gravitational wave observatories like LISA. In the limit presented here, for a static background potential and trajectories of the particle that are geodesic, it is straight-forward to compute the gravitational waveforms to illustrate how the zoom-whirl behaviour imprints itself. 

At lowest order, the gravitational waves are quadrupolar in nature, given by the quadrupole formula first derived by Einstein \cite{GW_History, einstein1918}
\beq
     h_{jk}^{TT} = \frac{2}{r}{\ddot
   {\cal I}}_{jk}^{TT}(t - r)\ .
    \label{eqn.quadrupoleFormula}
\eeq
The gravitational-wave amplitude is dimensionless. To restore conventional SI units to evaluate this expression, multiply it by $G/c^{4}$. Here ${\cal I}_{jk}(t)$ is the \textit{transverse traceless quadrupole moment tensor},
\beq
    {\cal I}_{jk} = \int \rho({\vec x})\left[x_{j}x_{k} - \frac{1}{3}r^{2}\delta_{jk} \right] d^{3}x\ ,
    \label{eqn.quadrupoleTensorTT}
\eeq
related to the familiar moment of inertia tensor in classical mechanics that characterizes the distribution of mass in a system.

Here, the coordinate positions $x^{i} = x^{i}(t)$ are  the position along the orbit as the particle moves in the effective potential, as a function of time. These expressions for the gravitational waves can be evaluated in our problem by replacing the density function with point particles of the appropriate mass, $\rho(t,{\vec x}) = \sum_{i} m_{i} \delta(x - x_{i}(t))$.

In practice, the trajectories $x^{i}(t)$ along the orbit can only be written down in closed form for circular orbits, where the solutions are sinusoids, and the time derivatives in Equation~\ref{eqn.quadrupoleFormula} can be written down analytically. For more general orbits, where the solutions are ellipses with varying speed along the ellipse, the solutions can be found by solving the Kepler Equation. In the case of integrated geodesic trajectories around the black hole, the $x^{i}(t)$ will generally be numerical solutions, and the derivatives can be evaluated numerically to give the gravitational waveform.

\begin{figure}
    \centering
   \includegraphics[width=0.75 \columnwidth]{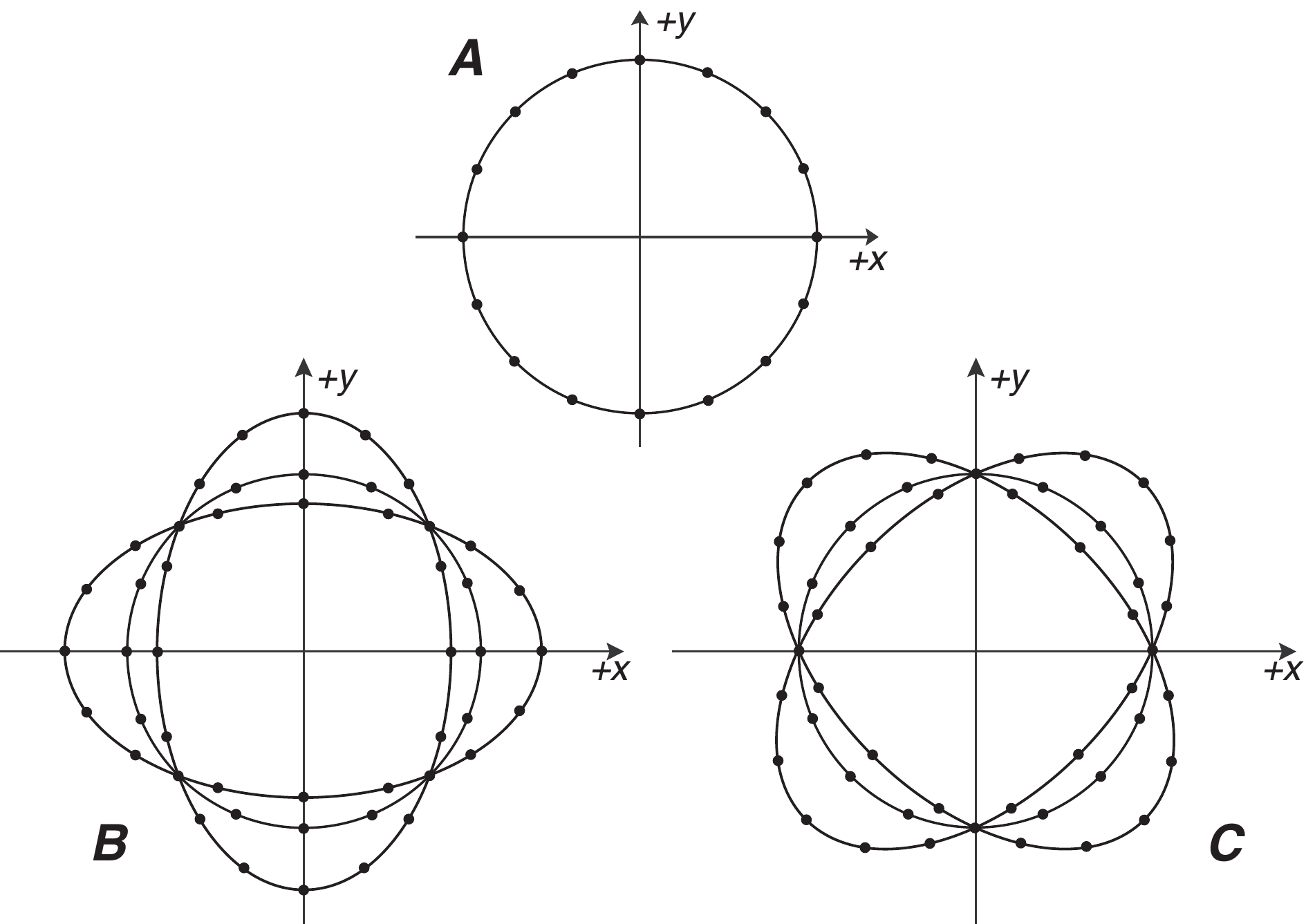}
   \caption{A circular ring of test particles in the $xy$-plane (panel A). The proper distance between particles in the ring is distorted, elongated along one axis and compressed along the orthogonal axis, oscillating as a gravitational wave propagates perpendicular to the ring. The case is shown for a $+$ polarized wave (panel B) and a $\times$ polarized wave (panel C).
   }
   \label{fig.ringTestParticles}
\end{figure}

Gravitational waves have two distinct polarization states, defined by the tidal distortion they produce when passing through a region of space. This distortion is \textit{transverse} to the direction of wave propagation, and is commonly envisioned as the time-dependent distortion of the proper distance across a ring of test particles, as shown in Figure~\ref{fig.ringTestParticles}. These polarization states are named ``plus'' (given by amplitude $h_{+}$) and ``cross'' (given by amplitude $h_{\times}$) after the principle axes the distortions occur along. This kind of naming will be familiar from electromagnetism, where vertical and horizontal polarization states are named by the axes along which stationary test charges feel a deflection force as an electromagnetic wave passes by. More complete discussions of the nature of gravitational waves, and the derivation of their polarization states and measuring their effects on matter, are covered in standard textbooks on general relativity. \cite{Moore, GWenv}

\begin{figure}
    \centering
   \includegraphics[width=0.75 \columnwidth]{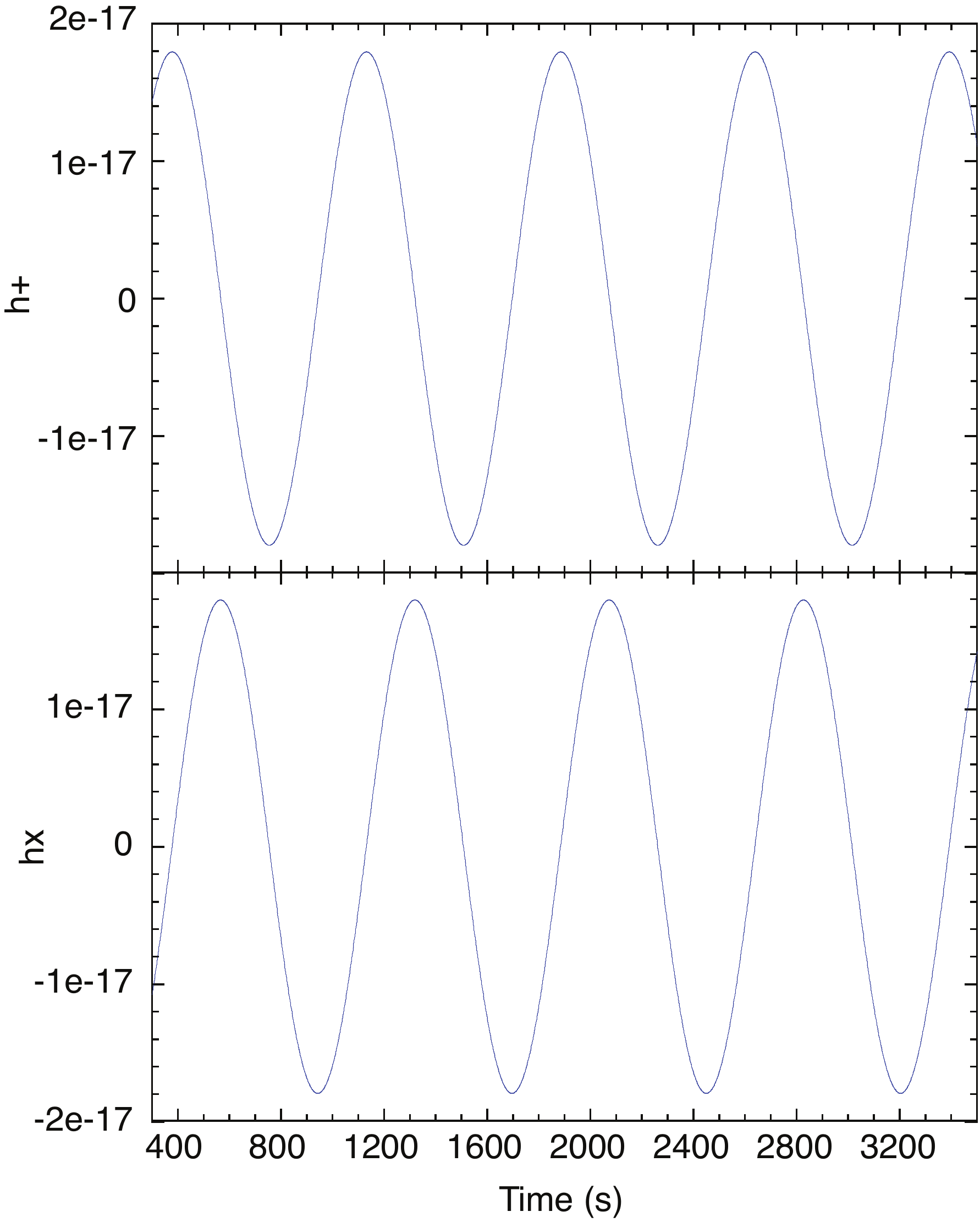}
   \caption{The $+$ (top panel) and $\times$ (bottom panel) waveforms for a circular ($e = 0.0$) orbit.
   }
   \label{fig.wavesCircular}
\end{figure}

\begin{figure}
    \centering
   \includegraphics[width=0.75 \columnwidth]{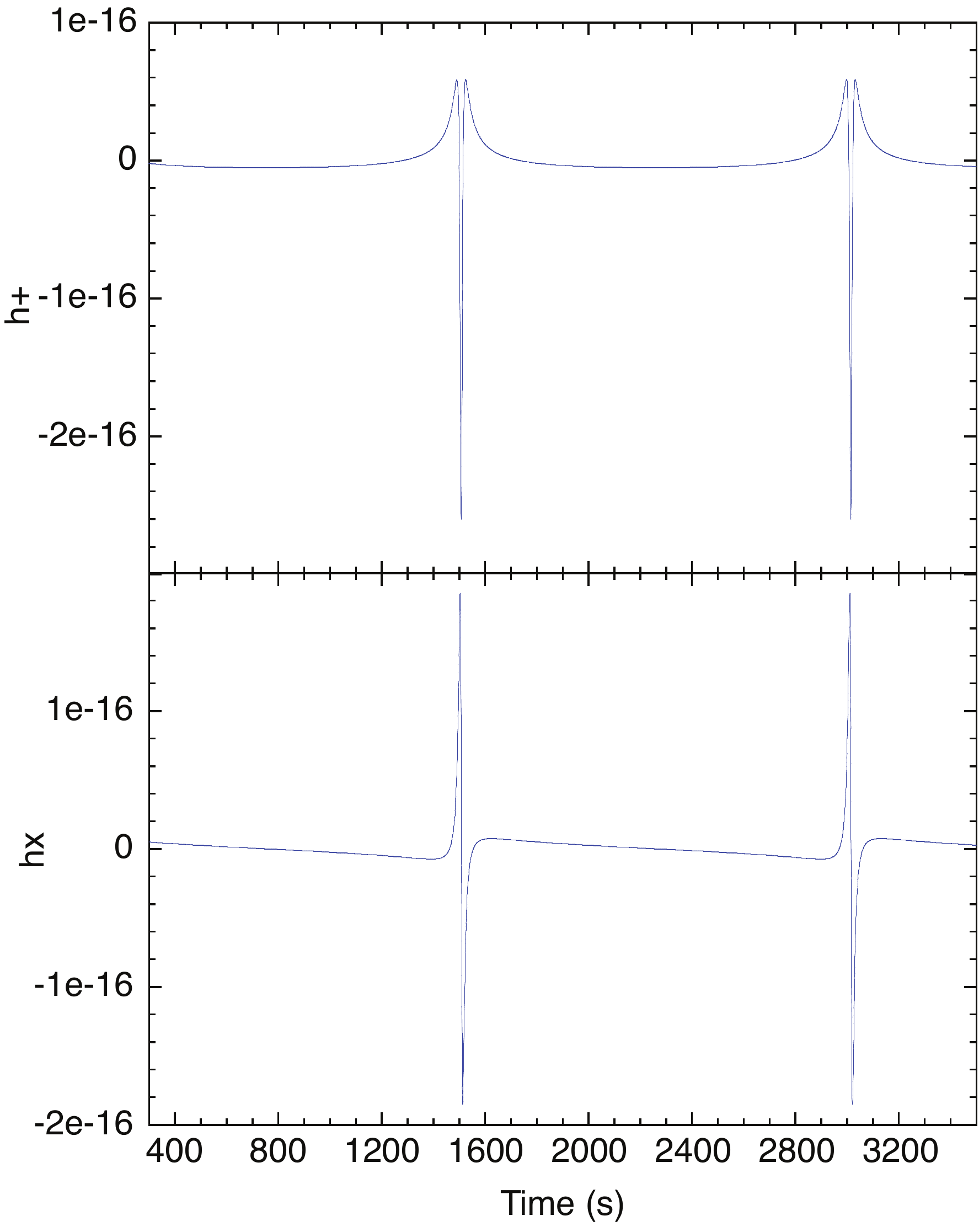}
   \caption{The $+$ (top panel) and $\times$ (bottom panel) waveforms for an eccentric ($e = 0.9$) orbit.
   }
   \label{fig.wavesEccentric}
\end{figure}

For circular orbits (at the minimum of the effective potential), the gravitational waveforms are perfect sinusoids, as shown in Figure~\ref{fig.wavesCircular}. As an orbit becomes more eccentric, the gravitational waveforms deviate from sinusoidal shapes, but remain periodic. Examples for orbits with moderate eccentricities are shown in Figure~\ref{fig.wavesEccentric} ($e = 0.9$). The waveform shows the highest amplitudes near periapsis, where the sources are moving most rapidly and the gravitational interactions are strongest (the source is ``more relativistic''). 

\begin{figure}
    \centering
   \includegraphics[width=0.8 \columnwidth]{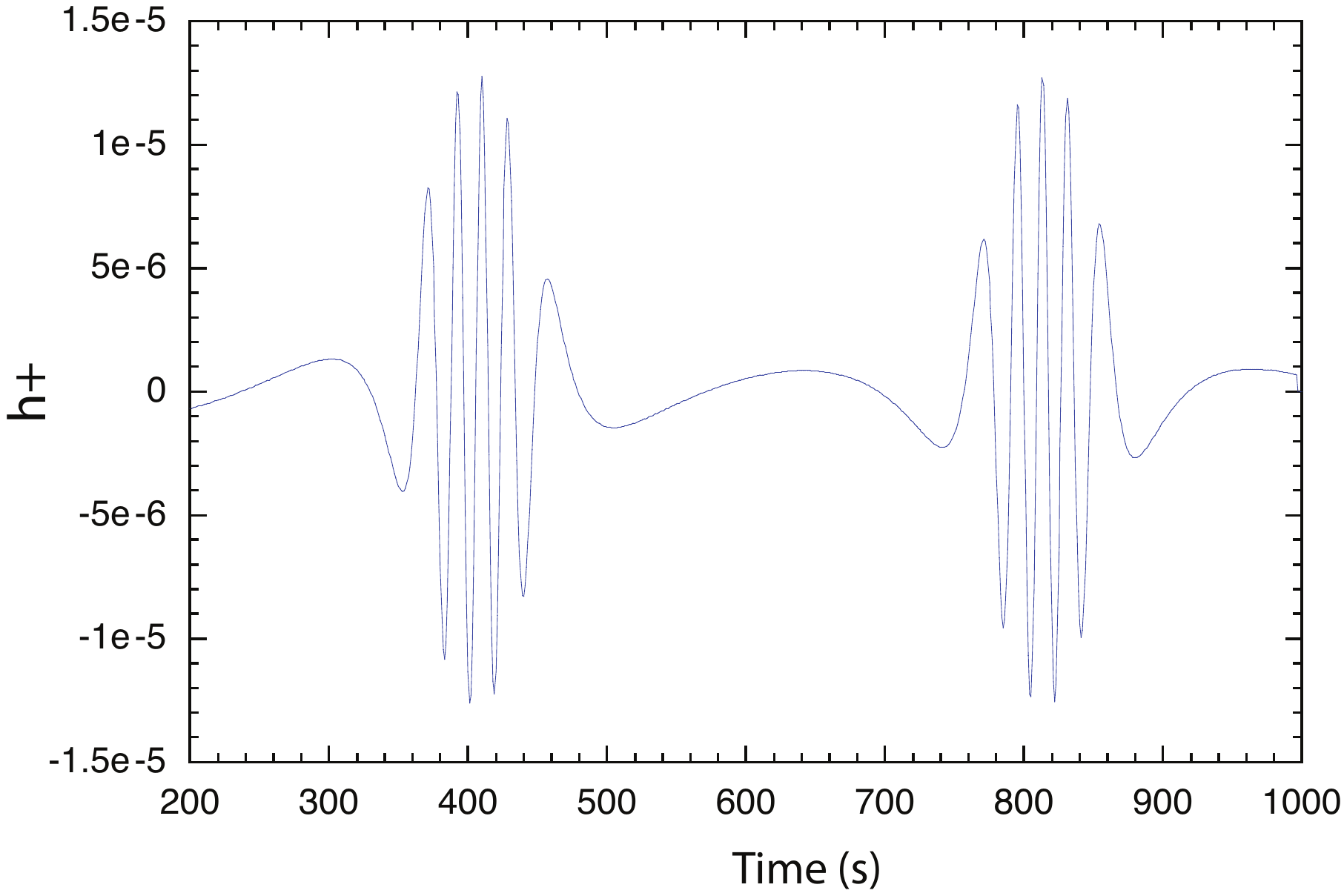}
   \caption{The $+$ waveform for an orbit in the whirling phase, for orbital energies near the peak of the effective potential (output waveform from the Dashboard \cite{zoomWhirlGithub}).
   }
   \label{fig.waveZoomWhirly}
\end{figure}

For extreme ``zoom-whirl'' orbits, of the sort EMRIs are expected to have, where the eccentricity is high and the periapsis (innermost turning point) is near the maximum in the effective potential, the waveforms have a very bursty appearance, but still periodic at the orbital period, as shown in Figure~\ref{fig.waveZoomWhirly}. The amplitude is high and the structure is complex while the particle is in the whirling phase, moving rapidly in close proximity to the central mass; when the particle zooms out to large distances, the interaction is much weaker and the gravitational-wave amplitude is comparatively tiny.  

%% ======= section : SIMULATION TOOL ================
\section{Simulation Tool} \label{sec.tool}

To facilitate exploration of using effectve potentials to understand complex gravitational-wave sources like EMRIs, we have developed a Simulation Dashboard that allows a user to change the parameters that define an orbit and/or the effective potential, then see a simultaneous visualization of the effective potential, the orbital trajectory, and the associated gravitational waveforms. The ``dashboard tool'' is provided as an open python package that can be downloaded and run locally (available on github, \cite{zoomWhirlGithub}). An online version that runs in a web-browser can be found at \href{http://ciera.northwestern.edu/gallery/zoom-whirl/}{ciera.northwestern.edu/gallery/zoom-whirl}.

\begin{figure}
    \centering
   \includegraphics[width=1
   \columnwidth]{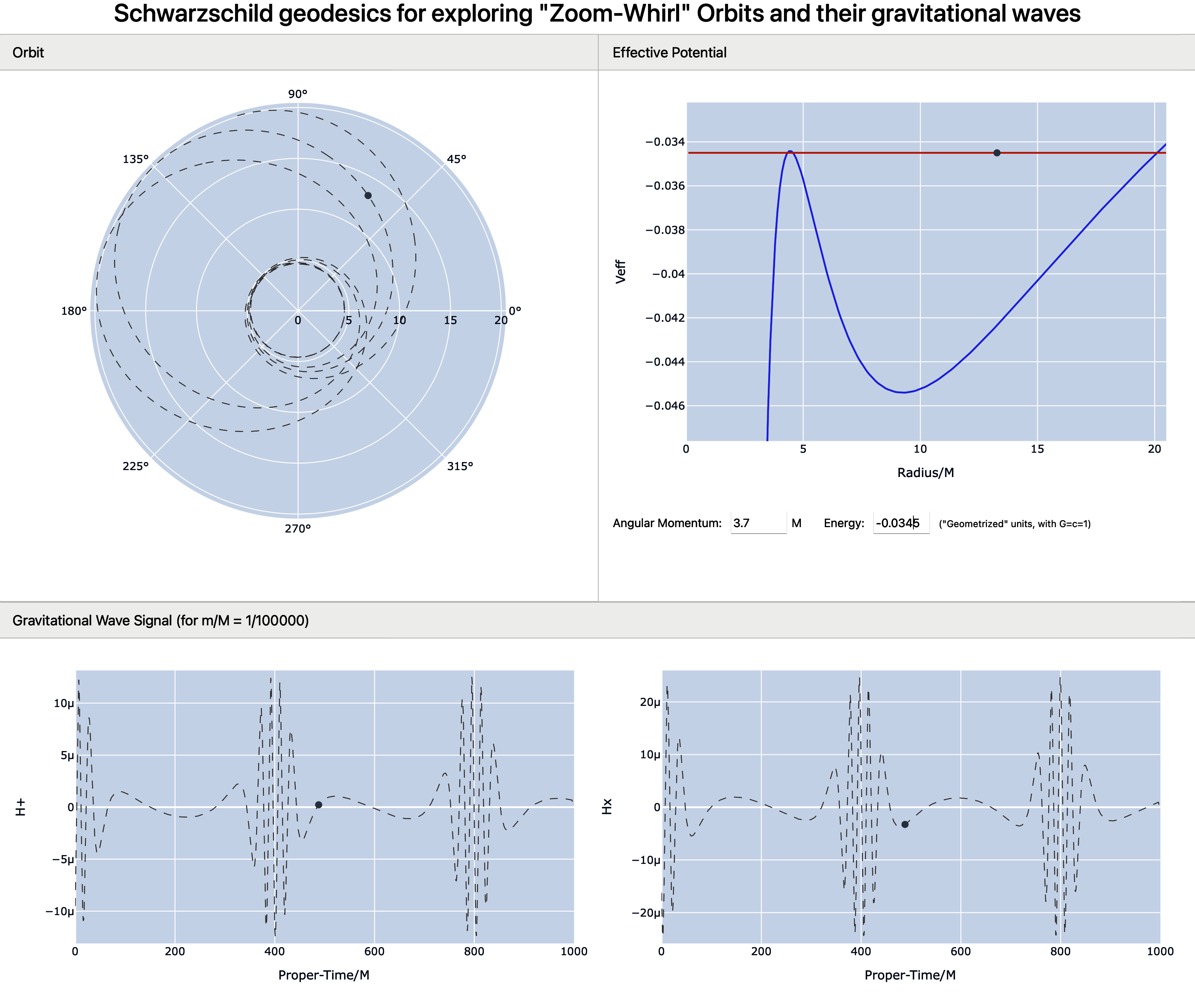}
   \caption{A screencapture of the Dashboard\cite{zoomWhirlGithub} in action, showing the trajectory of a zoom-whirl orbit (upper left), the particle in the effective potential (upper right), and the corresponding gravitational waveforms (lower panels).
   }
   \label{fig.dash_zoomWhirl}
\end{figure}

The tool presents a multi-pane dashboard, shown in Figure \ref{fig.dash_zoomWhirl} (Schwarzschild zoom-whirl). The appearance of the panes are controlled by user-configurable parameters that define the geodesic orbit of interest, each of which is uniquely identified by two constants of integration (either spatial parameters such as $r_p$ and $e$, or the traditional constants of integration ${\tilde E}$ and ${\tilde \ell}$).

The panes show the traditional effective potential plot with the energy of the current orbit displayed and the particle's radial location, an orbital trace looking down on the x-y plane from above with the particle's current radial location, and the two gravitational waveform polarizations, $h_+(t)$ and $h_{\times}(t)$. By adjusting the parameters that define the orbit, a user can in real-time observe how the turning points of the orbits evolve on the potential, and the impact on both the orbital trajectory and the gravitational waveforms.

%% ================== DISCUSSION ==========================
%% ========================================================
\section{Discussion}\label{sec.discussion}
In this paper, we have outlined how the seemingly exotic ``whirl-zoom'' behavior of EMRIs can be understood in the context of the shape of the effective potential the particles are moving in.  In the case of orbits in a Schwarzschild effective potential, conventional circular and elliptical orbits familiar from Newtonian theory can be found in the well near the local mimimum of the effective potential. However, for orbital parameters that give particle orbital energies that approach the inner local maximum of the potential, the familiar ``perihelion precession'' from relativity theory results. In the limit where the inner turning point of the orbit is near the peak of the effective potential, the perihelion precession becomes extreme, giving the ''whirling'' behaviour of EMRI orbits. 

We have captured these principles in a coded tool that can be run in a python environment or in a web-browser. Pedagogically, this is helpful for illustrating the utility of effective potentials in understanding orbital motion in a dynamic, interactive way, and provides a way to explore how traditional elliptical orbits are part of a continuous family of orbits that are related to the ``exotic'' zoom-whirl orbits from gravitational wave astronomy. This can be used in real-time to show the impact choices of effective potential have on particle motion and gravitational waveforms, but also can provide a kind of numerical laboratory for students to use in tandem with other analytic exercises they may be given as part of their regular classroom work.

With the code provided as open-source, we imagine it can be used in real-time in lecture presentations, but could also be useful foundation for exploratory homework problems or laboratory exercises in physics and astronomy courses.

\bibliography{zoomWhirl}  %% bibfile: zoomWhirl.bib

\end{document}